\documentclass[11pt,twoside]{article}
\usepackage{./asp2014}

\aspSuppressVolSlug
\resetcounters

\begin{document}

\title{Close binary progenitors and ejected companions of thermonuclear supernovae}
\author{S. Geier,$^1$ T. Kupfer,$^2$ U. Heber,$^3$ P. Nemeth,$^3$ E. Ziegerer,$^3$ A. Irrgang,$^3$ 
M. Schindewolf,$^3$ T. R. Marsh,$^4$ B. T. G\"ansicke,$^4$ B. N. Barlow,$^5$ S. Bloemen,$^6$}
\affil{$^1$Institute for Astronomy and Astrophysics, Kepler Center for Astro and Particle Physics, Eberhard Karls University, T\"ubingen, Germany; \email{geier@astro.uni-tuebingen.de}}
\affil{$^2$Division of Physics, Mathematics, and Astronomy, California Institute of Technology, Pasadena, USA}
\affil{$^3$Dr.~Karl~Remeis-Observatory \& ECAP, Astronomical Institute, Friedrich-Alexander University Erlangen-Nuremberg, Bamberg, Germany}
\affil{$^4$Department of Physics, University of Warwick, Conventry, UK}
\affil{$^5$Department of Physics, High Point University, High Point, USA}
\affil{$^6$Department of Astrophysics/IMAPP, Radboud University Nijmegen, Nijmegen, The Netherlands}

\paperauthor{S. Geier}{geier@astro.uni-tuebingen.de}{}{Eberhard Karls University}{Institute for Astronomy and Astrophysics, Kepler Center for Astro and Particle Physics}{T\"ubingen}{}{72074}{Germany}
\paperauthor{T. Kupfer}{tkupfer@caltech.edu}{}{California Institute of Technology}{Division of Physics, Mathematics, and Astronomy}{Pasadena}{CA}{91125}{USA}
\paperauthor{U. Heber}{heber@sternwarte.uni-erlangen.de}{}{Friedrich-Alexander University Erlangen-Nuremberg}{Dr.~Karl~Remeis-Observatory \& ECAP}{Bamberg}{}{96049}{Germany}
\paperauthor{P. Nemeth}{peter.nemeth@sternwarte.uni-erlangen.de}{}{Friedrich-Alexander University Erlangen-Nuremberg}{Dr.~Karl~Remeis-Observatory \& ECAP}{Bamberg}{}{96049}{Germany}
\paperauthor{E. Ziegerer}{eva.ziegerer@sternwarte.uni-erlangen.de}{}{Friedrich-Alexander University Erlangen-Nuremberg}{Dr.~Karl~Remeis-Observatory \& ECAP}{Bamberg}{}{96049}{Germany}
\paperauthor{A. Irrgang}{andreas.irrgang@sternwarte.uni-erlangen.de}{}{Friedrich-Alexander University Erlangen-Nuremberg}{Dr.~Karl~Remeis-Observatory \& ECAP}{Bamberg}{}{96049}{Germany}
\paperauthor{M. Schindewolf}{markus.schindewolf@fau.de}{}{Friedrich-Alexander University Erlangen-Nuremberg}{Dr.~Karl~Remeis-Observatory \& ECAP}{Bamberg}{}{96049}{Germany}
\paperauthor{T. R. Marsh}{t.r.marsh@warwick.ac.uk}{}{University of Warwick}{Department of Physics}{Coventry}{}{CV4 7AL}{UK}
\paperauthor{B. T. G\"ansicke}{Boris.Gaensicke@warwick.ac.uk}{}{University of Warwick}{Department of Physics}{Coventry}{}{CV4 7AL}{UK}
\paperauthor{B. N. Barlow}{bbarlow@highpoint.edu}{}{High Point University}{Department of Physics}{High Point}{NC}{27268}{USA}
\paperauthor{S. Bloemen}{s.bloemen@astro.ru.nl}{}{Radboud University Nijmegen}{Department of Astrophysics/IMAPP}{Nijmegen}{}{6500}{The Netherlands}

\begin{abstract}
Hot subdwarf stars (sdO/Bs) are evolved core helium-burning stars with very thin hydrogen envelopes, which can be formed by common envelope ejection. Close sdB binaries with massive white dwarf (WD) companions are potential progenitors of thermonuclear supernovae type Ia (SN~Ia). We discovered such a progenitor candidate as well as a candidate for a surviving companion star, which escapes from the Galaxy. More candidates for both types of objects have been found by crossmatching known sdB stars with proper motion and light curve catalogues. We found 72 sdO/B candidates with high Galactic restframe velocities, 12 of them might be unbound to our Galaxy. Furthermore, we discovered the second-most compact sdB+WD binary known. However, due to the low mass of the WD companion, it is unlikely to be a SN\,Ia progenitor.
\end{abstract}

\section{Introduction}

Hot subdwarf stars (sdO/Bs) are evolved core helium-burning stars with very thin hydrogen envelopes \citep{heber16}, which can be formed by common envelope ejection. Close sdB binaries with massive C/O-WD companions are candidates for supernova type Ia (SN~Ia) progenitors, because mass-transfer can lead to the thermonuclear explosion of the WD. The project Massive Unseen Companions to Hot Faint Underluminous Stars from SDSS (MUCHFUSS) aims at finding the sdB binaries with the most massive compact companions like massive white dwarfs, neutron stars or black holes.

We selected and classified about $\sim1400$ hot subdwarf stars from the Sloan Digital Sky Survey (SDSS DR7). Stars with high velocity variations have been reobserved and analysed. In total $177$ radial velocity variable subdwarfs have been dis\-covered and $1914$ individual radial velocities measured. We constrain the fraction of close massive companions of H-rich hot subdwarfs to be smaller than $\sim1.3\%$ \citep{geier15b}. Orbital parameters as well as minimum companion masses have been derived from the radial velocity curves of 30 sdB binaries \citep{kupfer15}. 

We detected high RV-variability of the bright sdB CD$-$30$^\circ$11223. Photometric follow-up revealed both shallow transits and eclipses, allowing us to determine its component masses and fundamental parameters. The binary system, which is composed of a C/O-WD ($\sim0.76\,M_{\rm \odot}$) and an sdB ($\sim0.51\,M_{\rm \odot}$) has a very short orbital period of $\sim0.049\,{\rm d}$. In the future mass will be transfered from the helium star to the white dwarf. After a critical amount of helium is deposited on the surface of the white dwarf, the helium is ignited. Modelling this process shows that the detonation in the accreted helium layer should be sufficiently strong to trigger the explosion of the core. Thermonuclear supernovae have been proposed to originate from this so-called double-detonation of a WD \citep{geier13}. 

The surviving companion star will then be ejected with its orbital velocity. The properties of such a remnant match the hypervelocity star US\,708, a helium-rich sdO star moving with $\sim1200\,{\rm km\,s^{-1}}$, exceeding the escape velocity of our Galaxy by far and making it the fastest unbound star known in our Galaxy \citep{geier15a}.

\articlefigure{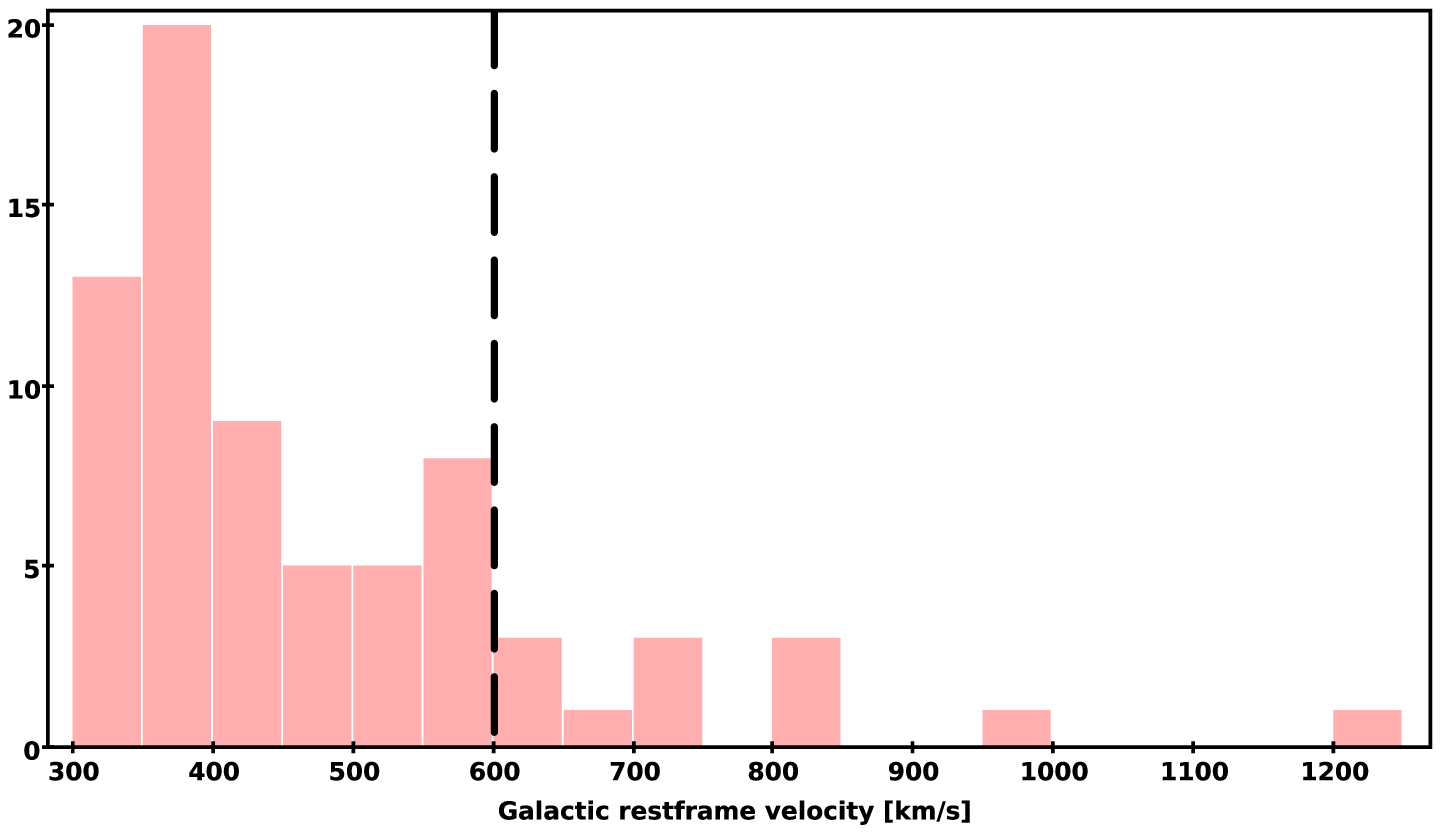}{vgrf}{Distribution of Galactic restframe velocities for our sample of 72 high-velocity sdO/B. The dashed vertical line marks the approximate escape velocity of the Galaxy.}

\section{Searching for ejected companion candidates}

Since the properties of the ejected companions, especially the ejection velocity, allow us to constrain the properties of the binary progenitors such as the orbital period and the companion mass right at the moment of the explosion \citep{geier15a}, we will gain an unprecedented insight into the formation of SN\,Ia and learn about other acceleration mechanisms for hypervelocity stars, if more such objects can be found and studied. The distribution of orbital periods and WD companion masses of progenitor binaries will help us to constrain SN\,Ia progenitor models. While binaries with periods longer than about 2\,hr will merge as double degenerates, closer binaries might be progenitors for the helium double-detonation channel.

To search for ejected companions we compiled a catalogue of essentially all known sdO/B stars from the literature and our own database ($\sim4900$ stars, Geier et al. in prep.) and crossmatched it with proper motion catalogues. Candidates with high Galactic restframe velocities are followed-up with spectroscopy (Keck/ESI, VLT/XSHOOTER, SOAR/Goodman, CAHA/TWIN, WHT/ISIS) to measure spectroscopic distances and derive kinematics. We found 72 sdO/B stars with preliminary Galactic restframe velocities exceeding $300\,{\rm km\,s^{-1}}$ and $12$ candidates for unbound hypervelocity sdO/Bs (see Fig.~\ref{vgrf}).

\section{Searching for close binary progenitor candidates}

Hot subdwarf binaries with massive WDs in close orbits turned out to be quite rare. To find more of those objects, we crossmatched the hot subdwarf catalogue with light curve catalogues (e.g. CRTS, PTF, SWASP) and search for the characteristic sinusoidal variations caused by the ellipsoidal deformation of the sdB. 

In course of this project we discovered the second-most compact sdB binary known \citep{kupfer16}. PTF1\,J082340.04+081936.5 has an orbital period as short as $0.060761584(10)\,{\rm d}$ and clearly shows light curve variations caused by an ellipsoidal deformation of the sdB primary (see Fig.~\ref{lc}) very similar to CD$-$30$^\circ$11223. However, the radial velocity semi-amplitude of this binary is significantly smaller ($K=211.7\pm1.8\,{\rm km\,s^{-1}}$, see Fig.~\ref{rv}) and the compact, invisible companion therefore less massive. 

Performing a combined analysis of the light curve and the time resolved spectroscopy of PTF1\,J082340.04+081936.5 we can constrain the mass of the sdB primary to $0.45_{-0.07}^{+0.09}\,M_{\rm \odot}$, which is quite typical, and the mass of the WD companion to $0.46_{-0.09}^{+0.12}\,M_{\rm \odot}$, close to the lower mass limit of C/O-WDs. Since the mass of the WD companion is too small for the double-detonation channel, which requires the WD to be more massive than $\sim0.8\,M_{\rm \odot}$), this system will either merge or evolve to become an AM\,CVn. 

\articlefigure{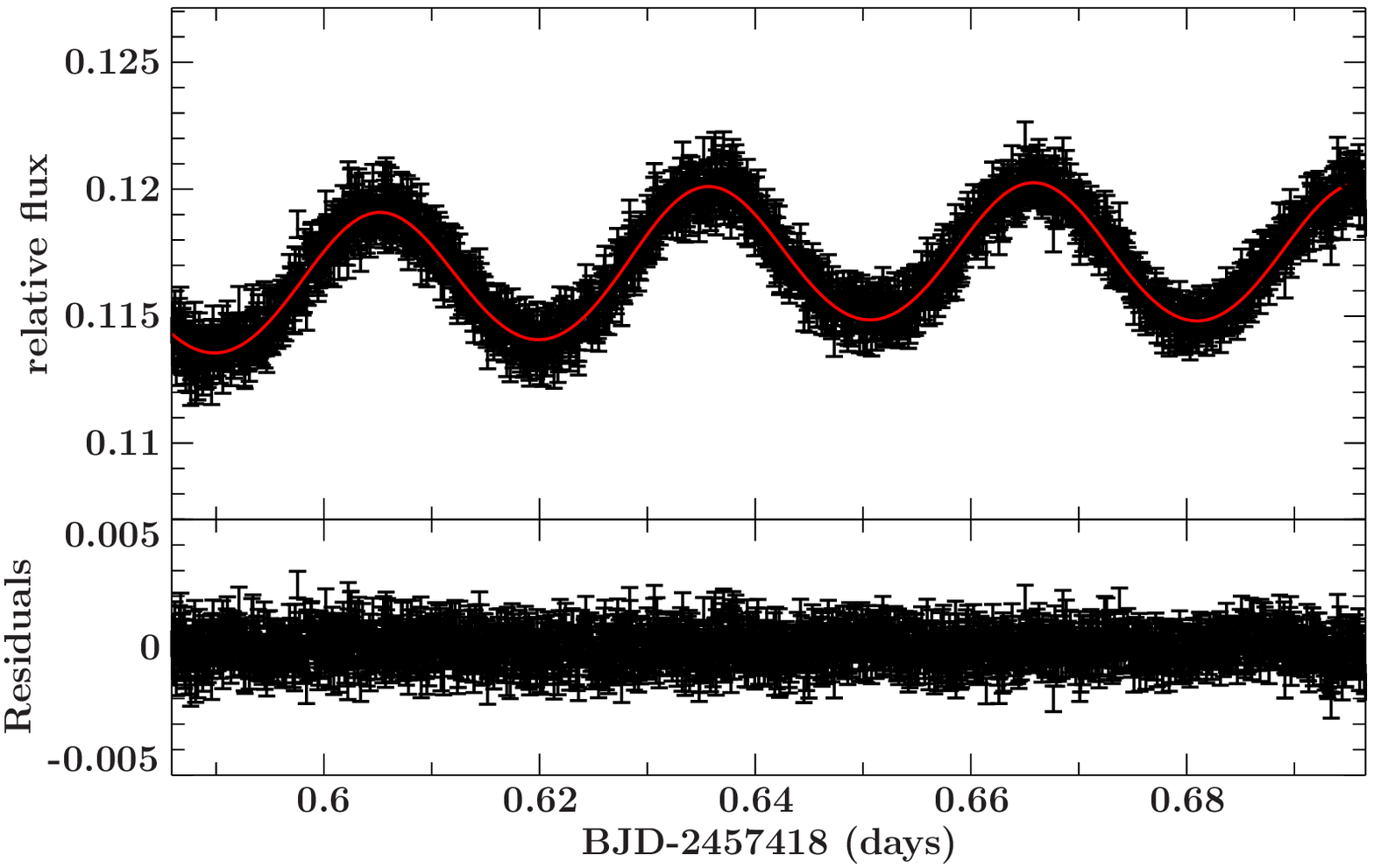}{lc}{Light curve of PTF1\,J082340.04+081936.5 shown with a model fit For details see \citet{kupfer16}.}

\articlefigure{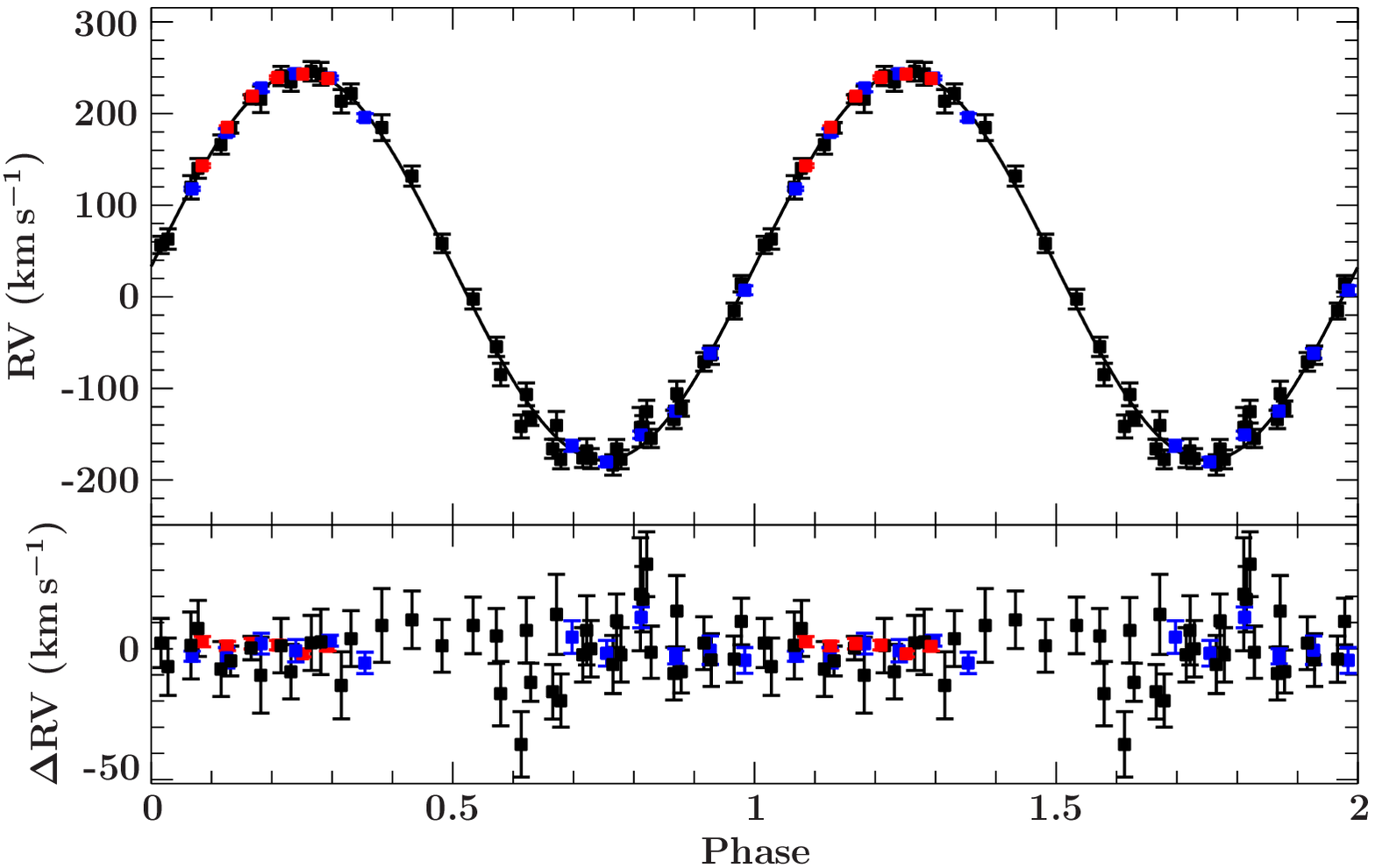}{rv}{Radial velocity curve of PTF1\,J082340.04+081936.5. For details see \citet{kupfer16}.}

The Gaia mission will provide accurate astrometry and light curves of all the stars in our hot subdwarf sample and will allow us to compile a much larger all-sky catalogue of those stars. Ongoing and upcoming ground-based surveys like ZTF, PanSTARRS, BlackGEM, NGTS and eventually LSST will add even more light curve information. In this way we expect to find many more progenitor binaries and ejected companions.

\clearpage 



\end{document}